\documentclass[a4paper,12pt]{article}

\usepackage{graphicx, amsmath, amsfonts, amssymb, array, natbib, graphicx, float, lipsum, pdflscape}

\title{Accounting for selection bias due to death in estimating the effect of wealth shock on cognition for the Health and Retirement Study}
\author{Yaoyuan V. Tan\footnote{(e-mail: yvt4@sph.rutgers.edu)} \\ Department of Biostatistics and Epidemiology, Rutgers University \\ and \\ Carol A.C. Flannagan \\\ Transportation Research Institute, University of Michigan \\ Lindsay R. Pool \\ Department of Preventive Medicine, Northwestern University\\ and \\ Michael R. Elliott \\ Department of Biostatistics, University of Michigan}
\date{}

\begin{document}
\maketitle
\linespread{2}
\selectfont

\begin{abstract}
The Health and Retirement Study is a longitudinal study of US adults enrolled at age 50 and older. We were interested in investigating the effect of a sudden large decline in wealth on the cognitive score of subjects. Our analysis was complicated by the lack of randomization, confounding by indication, and a substantial fraction of the sample and population will die during follow-up leading to some of our outcomes being censored. Common methods to handle these problems for example marginal structural models, may not be appropriate because it upweights subjects who are more likely to die to obtain a population that over time resembles that would have been obtained in the absence of death. We propose a refined approach by comparing the treatment effect among subjects who would survive under both sets of treatment regimes being considered. We do so by viewing this as a large missing data problem and impute the survival status and outcomes of the counterfactual. To improve the robustness of our imputation, we used a modified version of the penalized spline of propensity methods in treatment comparisons approach. We found that our proposed method worked well in various simulation scenarios and our data analysis.
\end{abstract}


\section{Introduction}
Late middle age adults commonly experience chronic health conditions like high blood pressure or diabetes as well as declining cognitive abilities. Factors known to be associated with accelerated decrease in cognitive abilities include smoking, high alcohol consumption, physical inactivity, high dietary intake of sodium and saturated fats, low dietary intake of fruits and vegetables \citep{lee_back,stuck}; hypertension, elevated serum cholesterol, diabetes, obesity, cerebrovascular and cardiovascular disease \citep{plassman}; depression, lower socioeconomic status, and exposure to acute stressful life events and chronic perceived stress \citep{krieger}. In particular, the acute stress of a sudden decrease in wealth -- ``a negative wealth shock'' -- may have a negative impact on the cognitive ability of late middle aged adults. Because income typically exceeds consumption at this stage in life, sudden decreases in wealth during this period not only decrease the amount of wealth saved for retirement, but there are fewer remaining years left to replenish the lost wealth \citep{butrica}. The stress of losing substantial wealth during the savings period of the life cycle coupled with the pressure to replenish the lost wealth can lead to stress-related health conditions which in turn reduces the cognitive ability of an individual \citep{shrira}. In addition, individuals who have received a negative wealth shock may have to reduce consumption of health-enhancing goods and services which in turn leads to poor management of existing chronic conditions, further reducing cognitive abilities \citep{friedman_con}.

Three issues arise when trying to estimate the causal effect of a negative wealth shock on cognitive ability. The first of these is the lack of randomization: negative wealth shocks are not randomly distributed in the population, but rather are confounded by factors such as gender and socio-economic status. The second issue is confounding by indication: the risk of the wealth shock at any point in time may depend on the prior cognitive ability up to the point. Finally, a sufficiently large fraction of the sample and the population will die during our follow-up, leading to ``censoring by death''. Those observed to have survived a negative wealth shock include those who would survive under either condition together with those that would survive only if they experienced a negative wealth shock (if any), while those observed to have survived in the absence of a negative wealth shock include those that would survive under either condition together with those that would survive only in the absence of a negative wealth shock. These ``missing values'' associated with cognition among the deceased are different from the measure of cognition being ``missing'' due to dropout, where the cognitive ability measure exists but is unobserved. As with wealth shock, death is not a random occurrence, and is positively associated with demographic measures that increase the risk of a negative wealth shock, increased cognitive ability decline, and the experience of a negative wealth shock. Hence, the measure for cognitive ability may be confounded by death if not considered appropriately. 

Methods have been developed to deal with these barriers to causal inference. To deal with the lack of randomization, we might hope that, conditional on available covariates, negative wealth shocks would truly be random. In this case, conditioning on the probability of receiving a negative wealth shock as a function of these covariates -- the propensity scores \citep{rosenbaum} -- can be used to remove the effect of confounding, either by regression, matching, or weighting \citep{imbens_rubin}. For the second issue -- confounding by indication -- marginal structural models \citep[MSM,][]{msm} and more recently, penalized spline of propensity methods in treatment comparisons \citep[PENCOMP,][]{zhou_t}, have been used to account for confounding by the time-dependence association of the cognitive measures, either by weighting using the inverse probability of treatment actually received based on the previous values of the time-varying covariates and outcomes (MSM), or by imputation of the missing counterfactual values (PENCOMP). For censoring by death, MSMs have typically been extended by multiplying the treatment assignment weights with the inverse of the predicted probability of death \citep{weuve}. The issue with this approach -- perhaps under appreciated -- is that the resulting pseudo-population is not only balanced with respect to exposure ``assignment'', but also ``immortal'', in the sense that those more likely to die are upweighted so that the population over time resembles that would have been obtained in the absence of death up till time $t$ \citep{chaix}. This is arguably not a sensible population for inference, at least from a policy and public health perspective.

A more refined approach would be to compare the difference in the effect of negative wealth shock on cognitive ability among subjects who would have survived whether they experienced a negative wealth shock or not. This approach is consistent with the potential outcomes approach of \cite{neyman} and \cite{rubin_po}, which defines causal effects as the within-subject difference of an outcome at a particular time under different exposure or treatment regimen, averaged over the population. This idea is not new \citep{elliott_bio} and can be viewed as a specific example of the principal stratification (PS) method discussed in \cite{frangakis}. Our innovation here is to embed this in a longitudinal setting where confounding by indication is present. We view this as a large missing data problem where survival status and, among survivors, unobserved outcomes under a given treatment pattern, are imputed. We extend the method proposed in Example 3 of \cite{elliott}, which provides a Bayesian MSM approach to compare two treatments at two time points. This approach was further extended by PENCOMP in \cite{zhou_t} which, like augmented inverse probability weighting \citep[AIPWT,][]{rrz}, has a doubly-robust property in that if either the mean or propensity model is correctly specified, consistent estimates of the causal effect will be obtained. We modified PENCOMP slightly using Bayesian additive regression trees (BART), a flexible model to ease the burden of model specification by the researcher, and apply this to our proposed method.

We organize our paper as follows. We set up the framework for our problem, and provide a brief review of of MSM, PENCOMP, and Bayesian additive regression trees (BART) in Section 2. We develop our proposed method in Section 3. We then explore some of the empirical properties of our proposed method compared to a na\"{i}ve method and MSM using a simulation study in Section 4. Section 5 describes the HRS data and the results of our negative wealth shock analysis. Section 6 concludes with a discussion of the implication of our results as well as future work.
   
\section{Review of Relevant Methods}
\label{chap3:review}
\subsection{Setup and notation}
Let $V=\{V_1,V_2,\ldots,V_p\}$ be $p$ baseline covariates, $Z_t$ be the treatment allocation at time $t=1,\ldots,T$ where $Z_t=1$ indicates a subject receiving a negative wealth shock at $t$ and $Z_t=0$ indicates no negative wealth shock, and $W_t=\{W_{1t},W_{2t},\ldots,W_{qt}\}$ be $q$ covariates that may vary with time, but are unaffected by a given treatment regimen. For example, fixed covariates by definition would belong to this class. Let $Y_{Z_1,\ldots,Z_t}$ be the potential outcome under treatments $Z_1,\ldots,Z_t$ and $X_{Z_1,\ldots,Z_t}=\{X_{Z_1,\ldots,Z_t,1},X_{Z_1,\ldots,Z_t,2},\ldots,X_{Z_1,\ldots,Z_t,r}\}$ be the time-varying covariates affected by treatments $Z_1,\ldots,Z_t$. Similarly, we define the potential survival indicator $S_{Z_1,\ldots,Z_{t-1}}$, for survival at time $t$. The survival outcome at $t$ measures whether a subject would survive after being exposed to treatment $Z_1,\ldots,Z_{t-1}$; hence, the lagged notation for the potential survival outcome, $S_{Z_1,\ldots,Z_{t-1}}$. $v$, $z_t$, $w_t$, $y_{z_1,\ldots,z_t}$, $x_{z_1,\ldots,z_t}$, and $s_{z_1,\ldots,z_t}$ indicate the observed baseline, treatment allocation, time varying covariates unaffected by a given treatment regimen, outcome, time-varying covariates affected by a given treatment regime, and survival status variables respectively. As in \cite{pool}, we assume that a negative wealth shock is an ``absorbing state'' so that once a subject receives a negative wealth shock at time $t$, i.e. $Z_t=1$, the subject is ``forever'' shocked, i.e. $Z_{t+1}=\ldots=Z_T=1$. Note that this need not be the case for a more general set up where we could have $Z_t=0$ when $Z_j=1$ for any $j=1,\ldots,t-1$. In our context, the potential outcomes for time $t=2$ are then $Y_{Z_1=0,Z_2=0}=Y_{00}$, $Y_{Z_1=0,Z_2=1}=Y_{01}$, and $Y_{Z_1=1,Z_2=1}=Y_{11}$; similarly, $X_{Z_1=0,Z_2=0}=X_{00}$, $X_{Z_1=0,Z_2=1}=X_{01}$, and $X_{Z_1=1,Z_2=1}=X_{11}$ for time-varying covariates under the various treatment regimes; and $S_{Z_1=0}=S_0$, $S_{Z_1=1}=S_1$ for survival states. Subjects who die at time $t$ have structurally missing data for outcomes and covariates i.e., $S_0=0$ implies that $Y_{00}=Y_{01}=NA$ and $X_{00}=X_{01}=NA$, while $S_1=0$ implies that $Y_{11}=NA$ and $X_{11}=NA$, where `NA' indicates a structurally missing observation.

\subsection{Marginal structural model \label{chap3:msm}}
To estimate the causal effect for confounding by indication and censoring by death problems, MSM makes the following assumptions. First, MSM assumes that 
\begin{equation}
	\label{chap3:surv_mod}
	P(S_{z_1,\ldots,z_{t-1}}|z_1,\ldots,z_{t-1},y_{z_1},\ldots,y_{z_1,\ldots,z_{t-1}},x_{z_1},\ldots,x_{z_1,\ldots,z_{t-1}},w_1,\ldots,w_{t-1},v)>0.
\end{equation}
and
\begin{equation}
	\label{chap3:treat_mod}
	P(Z_t|z_1,\ldots,z_{t-1},y_{z_1},\ldots,y_{z_1,\ldots,z_{t-1}},x_{z_1},\ldots,x_{z_1,\ldots,z_{t-1}},w_1,\ldots,w_{t-1},v)>0
\end{equation}
for any $z_t$ i.e. the probability of survival under treatment profile $z_1,\ldots,z_{t-1}$ and the probability of treatment allocation for time $t$ is bounded away from 0. This is an extension of the standard positivity assumption to allow that at least some subjects will survive under a given treatment regimen. Second, MSM assumes that there is no interference between subjects i.e. the potential outcome of subject $i$, $Y_{i,Z_1,\ldots,Z_t}=Y_{i,z_1,\ldots,z_t}$, is independent of whatever treatment regimen subject $j$ is allocated to $i\neq j$. Third, MSM assumes no unmeasured confounding and sequential randomization condition 
\[
	Y_{Z_1,\ldots,Z_t}\bot Z_t|z_1,\ldots,z_{t-1},y_{z_1,\ldots,z_{t-1}},\ldots,y_{z_1},x_{z_1,\ldots,z_{t-1}},\ldots,x_{z_1},w_1,\ldots,w_{t-1},v.
\] 
Finally, MSM assumes that the model specifications for Equations \ref{chap3:surv_mod}, \ref{chap3:treat_mod}, and
\[
	Y_{z_1,\ldots,z_t}|z_1,\ldots,z_{t},y_{z_1,\ldots,z_{t-1}},\ldots,y_{z_1},x_{z_1,\ldots,z_{t-1}},\ldots,x_{z_1},w_1,\ldots,w_{t-1},v
\]
are correct.

With these assumptions in place, $E[Y_{z_1,\ldots,z_t}-Y_{z'_1,\ldots,z'_t}]$ (note that this estimand is not conditioned on the survival status) is obtained by maximizing the weighted likelihood of
\begin{equation}
	\prod_{i=1}^n f(Y_{i;z_1,\ldots,z_t}|\mathbf{\theta}_{it})^{w_{it}},
\end{equation}
where $i$ indexes the subjects and $\mathbf{\theta}_{it}$ are the parameters involved in the model for $Y_{i;z_1,\ldots,z_t}$ and
\begin{equation}
	\label{chap3:w_treat}
	\small
	w_{it}=[\prod_{j=1}^tP(Z_{ij}=z_{ij}|z_{i1},\ldots,z_{i,j-1},y_{i1},\ldots,y_{i,j-1},x_{i1},\ldots,x_{i,j-1},w_{i1},\ldots,w_{i,j-1},v_i;\mathbf{\tau}_j)]^{-1}.
\end{equation}
By weighting using the inverse probability of receiving the observed treatment regime given all covariates and previous treatments, the association between treatment and all observed confounders, including confounding by indication, are broken. Under these four assumptions, inference about the treatment effects under a pseudo-population in which treatment is randomized can then be obtained.

Similarly, this weighting method can be used to remove bias due to dropout. Let $R_i=1$ indicate that the subject's cognitive score is observed and $R_i=0$ indicate that the subject's cognitive score is missing. The weight used to account for missing cognitive score is
\begin{equation}
	\label{chap3:w_miss}
	\scriptsize
	w_{it}^r=[\prod_{j=1}^tP(R_{ij}=r_{ij}|r_{i1},\ldots,r_{i,j-1},z_{i1},\ldots,z_{i,j-1},y_{i1},\ldots,y_{i,j-1},x_{i1},\ldots,x_{i,j-1},w_{i1},\ldots,w_{i,j-1},v_i;\mathbf{\gamma}_j)]^{-1}.
\end{equation}

Finally, death is typically treated as equivalent to dropout in MSM \citep{do, pool}. Let $D_{it}=1$ indicate that subject $i$ is dead at time $t$ and $D_{it}=0$ indicate that the subject survived at time $t$ (thus $D_{it}=1-S_{it}$). The weight for death censoring is then
\begin{equation}
	\label{chap3:w_death}
	\small
	w_{it}^d=[\prod_{j=1}^tP(D_{ij}=d_{ij}|z_{i1},\ldots,z_{i,j-1},y_{i1},\ldots,y_{i,j-1},x_{i1},\ldots,x_{i,j-1},w_{i1},\ldots,w_{i,j-1},v_i;\mathbf{\lambda}_j)]^{-1}.
\end{equation}

Assuming that these three weights are independent of each other, the final weight that we used becomes $w_{it}^f=w_{it}w_{it}^dw_{it}^r$. To stabilize the weights, the numerators of Equations \ref{chap3:w_treat}, \ref{chap3:w_miss}, and \ref{chap3:w_death} are replaced by the marginal probabilities of treatment, dropout, and death at baseline given by
\[
	\prod_{j=1}^tP(Z_{ij}=z_{ij}|z_{i1},\ldots,z_{i,j-1},v_i;\mathbf{\tau}_j'),
\]
\[
	\prod_{j=1}^tP(R_{ij}=r_{ij}|r_{i1},\ldots,r_{i,j-1},v_i;\mathbf{\gamma}_j'),
\]
and
\[
	\prod_{j=1}^tP(D_{ij}=d_{ij}|v_i;\mathbf{\lambda}_j')
\]
respectively. We use the stabilized weights in our simulations and analysis.

\subsection{Penalized Spline of Propensity Methods for Treatment Comparison}
PENCOMP uses the same four assumptions made by MSM excluding Equation \ref{chap3:surv_mod} for confounding by indication problems. Full details of PENCOMP can be found in \cite{zhou_t}. We briefly describe the algorithm for PENCOMP using multiple imputation (MI) with longitudinal treatment assignments here. Without loss of generality, we assume no time-varying covariates in the data.
\begin{enumerate}
	\item For $b=1,\dots,B$, generate a bootstrap sample $S^{(b)}$ from the original data $S$ by sampling units with replacement, stratified on treatment group. For each sample $b$, carry out steps 2-7.
	\item Estimate a logistic regression model for the distribution of $Z_1$ given baseline covariates $V$ with regression parameters $\gamma_{z_1}$. Estimate the propensity to be assigned treatment $Z_1=z_1$ as $\hat{P}_{z_1}(V)=Pr(Z_1=z_1|V;\hat{\gamma}_{z_1}^b)$, where $\hat{\gamma}_{z_1}^b$ is the maximum likelihood (ML) estimate of $\gamma_{z_1}$. Define $\hat{P}^*_{z_1}=\log[\frac{\hat{P}_{z_1}(V)}{1-\hat{P}_{z_1}(V)}]$.
	\item Using the cases assigned to treatment group $Z_1=z_1$, estimate a normal linear regression of $Y_{z_1}$ on $V$, with mean 
	\begin{equation}
		\label{chap3:pencomp}
		E(Y_{z_1}|V,Z_1=z_1,\theta_{z_1},\beta_{z_1})=s(\hat{P}^*_{z_1}|\theta_{z_1})+g_{z_1}(\hat{P}^*_{z_1},V;\beta_{z_1}),
	\end{equation}
	where $s(\hat{P}*_{z_1}|\theta_{z_1})$ denotes a penalized spline with fixed knots and parameters $\theta_{z_1}$ and $g_{z_1}(.)$ represents a parametric function of other predictors of the outcome, indexed by parameters $\beta_{z_1}$. One of the covariates might be omitted to avoid collinearity in the covariates in Equation \ref{chap3:pencomp}.
	\item For $z_1=0,1$, impute the values of $Y_{z_1}$ for subjects in treatment group $1-z_1$ in the original data with draws from the predictive distribution of $Y_{z_1}$ given $V$ from the regression in Step 3, with the ML estimates $\hat{\theta}_{z_1}^{(b)},\hat{\beta}_{z_1}^{(b)}$ substituted for the parameters $\theta_{z_1}^{(b)},\beta_{z_1}^{(b)}$.
	\item Estimate a logistic regression model for the distribution of $Z_2$ given $V,Z_1,(Y_0,Y_1)$, with regression parameters $\gamma_{z_2}$ and missing values of $(Y_0,Y_1)$ imputed from Step 4. Estimate the propensity to be assigned treatment $Z_2=z_2$ given $Z_1$, $Y_{Z_1}$, and $V$ as $\hat{P}_{z_2}(Z_1,Y_{Z_1},V)=Pr(Z_2=z_2|Z_1=z_1,Y_{z_1},V;\hat{\gamma}_{z_2}^{(b)})$, where $\hat{\gamma}_{z_2}^{(b)}$ is the ML estimate of $\gamma_{z_2}$. The probability of treatment regimen $(Z_1=z_1,Z_2=z_2)$ is denoted as $\hat{P}_{z_1z_2}=\hat{P}_{z_1}(V)\hat{P}_{z_2}(Z_1,Y_{Z_1},V)$, and define $\hat{P}^*_{z_1,z_2}=\log[\frac{\hat{P}_{z_1z_2}}{1-\hat{P}_{z_1z_2}}]$.
	\item Using the cases assigned to treatment group $(z_1,z_2)$, estimate a normal linear regression of $Y_{z_1,z_2}$ on $Z_2$, $Z_1$, $Y_{Z_1}$, and $V$ with mean 
	{\scriptsize
	\begin{equation}
		\label{chap3:pencomp2}
		E(Y_{z_1,z_2}|V,Y_{z_1},Z_1=z_1,Z_1=z_2,\theta_{z_1,z_2},\beta_{z_1,z_2})=s(\hat{P}^*_{z_1,z_2}|\theta_{z_1,z_2})+g_{z_1,z_2}(\hat{P}^*_{z_1,z_2},Z_2,Z_1,Y_{Z_1},V;\beta_{z_1,z_2}).
	\end{equation}
	}
	\item For each combination of $(z_1,z_2)$ impute the values of $Y_{z_1,z_2}$ for subjects not assigned this treatment combination in the original data with draws from the predictive distribution of $Y_{z_1,z_2}$ in Step 6, with ML estimates $\hat{\theta}_{z_1,z_2}^{(b)},\hat{\beta}_{z_1,z_2}^{(b)}$ substituted for the parameters $\theta_{z_1,z_2}^{(b)},\beta_{z_1,z_2}^{(b)}$. Let $\hat{\Delta}_{01,00}^{(b)}=E[Y_{01}-Y_{00}]$, $\hat{\Delta}_{11,00}^{(b)}=E[Y_{11}-Y_{00}]$, and $\hat{\Delta}_{11,01}^{(b)}=E[Y_{11}-Y_{01}]$ denote the average treatment effects, $\hat{\Delta}_{jk,lm}^{(b)}$, with associated pooled variance estimates $W_{jk,lm}^{(b)}$, based on the observed and imputed values of $Y$ for each treatment regimen.
	\item The MI estimate of $\Delta_{jk,lm}$ is then $\bar{\Delta}_{jk,lm,B}=\sum_{b=1}^B\hat{\Delta}_{jk,lm}^{(b)}$, and the MI estimate of the variance of $\bar{\Delta}_{jk,lm}$ is $T_B=\bar{W}_{jk,lm,B}+(1+1/B)D_{jk,lm,B}$, where $\bar{W}_{jk,lm,B}=\sum_{b=1}^BW_{jk,lm}^{(b)}/B$, $D_{jk,lm,B}=\sum_{b=1}^B\frac{(\hat{\Delta}_{jk,lm}^{(b)}-\bar{\Delta}_{jk,lm,B})^2}{B-1}$. The estimate $\Delta_{jk,lm}$ follows a $t$ distribution with degree of freedom $\nu$, $\frac{\Delta_{jk,lm}-\bar{\Delta}_{jk,lm,B}}{\sqrt{T_B}}\sim t_{\nu}$, where $\nu=(B-1)(1+\frac{\bar{W}_{jk,lm,B}}{D_{jk,lm,B}(B+1)})^2$.
\end{enumerate}

\subsection{Bayesian additive regression trees}
BART \citep{chipman_bart} is a flexible estimation technique for any arbitrary function. Suppose we have a continuous outcome $Y$ and corresponding $p$ predictors $X=(X_1,\ldots,X_p)$. Suppose $Y$ is related to $X$ via
\begin{equation}
	Y=f(X)+e
\end{equation}
where $f(.)$ is any arbitrary function which could involve complicated non-linear and multiple-way interactions and $e\sim N(0,\sigma^2)$. Formally, BART is written as
\begin{equation}
	\label{chap3:bart_eq}
	Y=\sum_{j=1}^mg(X,T_j,M_j)+e
\end{equation}
where $(T_j,M_j)$ is the joint distribution of the $j^{\text{th}}$ binary tree structure $T_j$ with its corresponding $b_j$ terminal node parameters $M_j=(\mu_{1j},\ldots,\mu_{b_jj})$. $m$ is the number of regression trees used to estimate $f(X)$ and it is usually fixed at 200.

BART is able to model multiple-way interactions by using regression trees. In essence, a binary regression tree in BART may be viewed as a penalized form of an Analysis of Variance (ANOVA) model. When the binary regression tree only splits on one variable for the whole tree, a main effects model is obtained. When the regression tree involve splits on many different variables, a multiple-way interaction model is obtained. BART combines all $m$ regression trees together in an additive manner to obtain non-linear estimates of the main and interaction effects. This additive procedure is done by first `breaking' $Y$ into $m$ equal `pieces' and fitting a regression tree to each piece. Subsequently, the regression tree in each $m$ piece is then estimated by looking at the residual produced by the other $m-1$ most updated regression trees. MCMC procedures are then used to obtain the posterior distribution of $f(X)$. When the default priors of BART suggested by \cite{chipman_bart} are assumed, the MCMC ensures that the eventual distribution of the the sum of regression trees is concentrated around the true distribution of the model \citep{rockova}.

For binary outcomes, BART uses a probit link where
\begin{equation}
	\label{chap3:bin_bart_eq}
	P(Y=1|X)=\Phi(\sum_{j=1}^mg[X,T_j,M_j])
\end{equation} 
where $\Phi(.)$ is the cdf of a standard normal distribution. Estimation of the posterior distribution is similar to that of continuous outcomes but with the use of data augmentation methods, i.e. draw a continuous latent variable based on whether $Y=1$ or $Y=0$ and then run the BART algorithm on the drawn latent variables.

\cite{kapelner_miss} suggested a procedure to allow the BART algorithm to include covariates that might contain missing values. In brief, the missingness in the covariates are not imputed but instead, viewed as a `value level' in the MCMC algorithm. The MCMC algorithm then `sends' missing data to terminal nodes in the regression trees that would maximize the likelihood. This is termed as ``Missing Incorporated in Attributes'' \citep[MIA,][Section 2]{twala}. \cite{kapelner_miss} showed using simulation examples that incorporating MIA within BART allows the appropriate handling of different types of missing mechanism, MCAR, MAR, and NMAR, for each covariate. We utilize this approach to accommodate the missingness in our covariates for the data analysis.

\section{Dealing with Censoring by Death}
\label{chap3:method}
\subsection{Determining the principal strata}
To determine the principal strata definition, we first investigated what the data for our problem could potentially look like. We constructed Table \ref{chap3:eg2_1} for $t=3$, $p=1$, and no time-varying covariates without loss of generality. In this table, `x' indicates an observed value, `?' represent a missing observation which needs to be imputed, and `NA' indicates a structurally missing observation. For the potential survival outcomes, we did not indicate whether they were missing or observed because we wanted to use Table \ref{chap3:eg2_1} to help us decide how we should be stratifying our subjects once our proposed method imputes the counterfactual survival status.

From Table \ref{chap3:eg2_1}, we can see that the goal of our analysis is to provide inference about $E[Y_{Z_1,\ldots,Z_t}-Y_{Z'_1,\ldots,Z'_t}|S_{Z_1,\ldots,Z_{t-1}}=S_{Z'_1,\ldots,Z'_{t-1}}=1]$, where $Z_l\neq Z'_l$ for at least one $l$ with $l=1,\ldots,t$ i.e. we condition on subjects who would potentially survive under two different treatment regimes $Z_1,\ldots,Z_{t-1}$ and $Z'_1,\ldots,Z'_{t-1}$. Thus, the distribution of $(S_{Z_1,\ldots,Z_{t-1}},S_{Z'_1,\ldots,Z'_{t-1}})$ form our principal strata and meaningful contrasts are defined only in the stratum where $S_{Z_1,\ldots,Z_{t-1}}=S_{Z'_1,\ldots,Z'_{t-1}}=1$ since the potential outcomes for the two different treatment regimes exist only in this stratum. For example, if we want to estimate the effect for a negative wealth shock at $t=2$ versus no negative wealth shock by $t=2$ that is $E[Y_{01}-Y_{00}|S_0=1]$, we restrict to subjects who survive if they did not receive a negative wealth shock at $t=1$ i.e. subjects with $S_0=1$ (Subjects 1-12 in Table \ref{chap3:eg2_1}). Note that the definition, $E[Y_{Z_1,\ldots,Z_t}-Y_{Z'_1,\ldots,Z'_t}|S_{Z_1,\ldots,Z_{t-1}}=S_{Z'_1,\ldots,Z'_{t-1}}=1]$, is different from the parameter MSM estimates which is $E[Y_{Z_1,\ldots,Z_t}-Y_{Z'_1,\ldots,Z'_t}]$.

\subsection{Proposed method}
\label{chap3:prop_meth}
We make the same four assumptions used by MSM (See Section \ref{chap3:msm}). We impose a further fifth assumption of strict monotonicity in that 
\begin{equation}
	\label{monotonicity}
	\text{If}\,Z_1\leq Z'_1,Z_2\leq Z'_2,\ldots,Z_{t-1}\leq Z'_{t-1},\,Z_i\neq Z'_i\,\text{for any }i,\,\text{then}\,S_{Z_1,\ldots,Z_{t-1}}\geq S_{Z'_1,\ldots,Z'_{t-1}}.
\end{equation}
As a consequence, we have for example, when $t=2$, if $S_0=0$ then $S_1=0$ and if $S_1=1$ then $S_0=1$. This means that we rule out the possibility of a subject who does not receive a negative wealth shock and dies but would survive if having received a neagtive wealth shock. Conversely, if a subject survives after having received a negative wealth shock, we rule out the possibility that this same subject would die if he or she did not receive a negative wealth shock. 
 
Our proposed method then estimates $E[Y_{Z_1,\ldots,Z_t}-Y_{Z'_1,\ldots,Z'_t}|S_{Z_1,\ldots,Z_{t-1}}=S_{Z'_1,\ldots,Z'_{t-1}}=1]$ by imputing the survival status of each subject at the current time $t$ and then combine the imputed counterfactual survival status together with the observed survival status to determine which principal stratum a subject belongs to. We then use a slightly modified PENCOMP to impute the counterfactual outcomes among the potentially surviving subjects to account for the bias due to confounding by indication. This approach is doubly robust and reduces the burden of model specification by the researcher. Subsequently, the average difference in the treatment effect within the desired principal strata is calculated. Variance is estimated using Rubin's combine rule to account for the imputation uncertainty \citep{heitjan}. Detailed steps for our method are given below.
\begin{enumerate}
	\item Generate a bootstrap sample $b$ from the data by sampling the units with replacement.
	\item Estimate the model $X_{z_1^{(b)}}^{(b)}|Z_1^{(b)}=z_1^{(b)},W_1^{(b)},V^{(b)}$. Use this model to compute the counterfactual of $X_{z_1^{(b)}}^{(b)}$ for bootstrap sample $b$.
	\item Estimate the distribution of $Z_1^{(b)}|W_1^{(b)},V^{(b)}$. Use this model to estimate the propensity to be assigned treatment $Z_1^{(b)}=z_1^{(b)}$ as $P^*_{z_1^{(b)}}=Pr(Z_1^{(b)}=z_1^{(b)}|W_1^{(b)},V^{(b)})$. Note that we did not perform a logit transformation to obtain $P^*_{z_1^{(b)}}$ (See PENCOMP Steps 2 and 5). This is because by using PENCOMP modified with BART to predict the outcomes, the non-linear effect of the propensity of assigned treatment will be handled automatically. Hence, any non-linear transformation on the propensity of assigned treatment would not be needed. 
	\item Estimate the model $Y_{z_1^{(b)}}^{(b)}|P^*_{z_1^{(b)}},Z_1^{(b)}=z_1^{(b)},X_{z_1^{(b)}}^{(b)},W_1^{(b)},V^{(b)}$. As mentioned, we used PENCOMP modified with BART to estimate this model. The advantage of using BART is the researcher no longer needs to specify the model. BART automatically takes care of any linear or non-linear main effects as well as linear or non-linear interactions. If we observe Equations \ref{chap3:pencomp} and \ref{chap3:pencomp2}, we can see that these two equations are constructed using a non-linear spline specification on the propensity of assigned treatment combined with possible linear interactions between the propensity of assigned treatment and remaining covariates. This fits well with the type of estimation problems that BART was designed to solve. We then use the model produced by BART-modified PENCOMP to compute the counterfactual of $Y_{z_1^{(b)}}^{(b)}$ for bootstrap sample $b$.
	\item Estimate the distribution for $S_{z_1^{(b)}}^{(b)}|Z_1^{(b)}=z_1^{(b)},Y_{z_1^{(b)}}^{(b)},X_{z_1^{(b)}}^{(b)},W_1^{(b)},V^{(b)}$ at $t=2$. Use this model to generate a survival status for the counterfactual of $S_{z_1^b}^{(b)}$ taking into account the assumption of monotonicity in Equation (\ref{monotonicity}) i.e. if $S_0$ is observed and $S_0=0$ then $S_1=0$. Similarly, if $S_1$ is observed and $S_1=1$ then $S_0=1$.	
	\item Estimate the model $X_{z_1^{(b)},z_2^{(b)}}^{(b)}|Z_1^{(b)}=z_1^{(b)},Z_2^{(b)}=z_2^{(b)},Y_{z_1^{(b)}}^{(b)},X_{z_1^{(b)}}^{(b)},W_1^{(b)},W_2^{(b)},V^{(b)}$. Use the respective models to impute the counterfactual of $X_{z_1^{(b)},z_2^{(b)}}^{(b)}$, using any previously imputed values for the unobserved treatment regimes and restricting to the subjects that are observed and predicted to survive under the given treatment regimen of interest at $t=1$.
	\item Estimate the distribution of $Z_2^{(b)}|Z_1^{(b)}=z_1^{(b)},Y_{z_1^{(b)}}^{(b)},X_{z_1^{(b)}}^{(b)},W_1^{(b)},W_2^{(b)},V^{(b)}$. Use this model to estimate the propensity to be assigned treatment $Z_2^{(b)}=z_2^{(b)}$ as $P_{z_2^{(b)}}=Pr(Z_1^{(b)}=z_1^{(b)}|X_{z_1^{(b)}}^{(b)},Z_1^{(b)}=z_1^{(b)},W_1^{(b)},V^{(b)})$. The probability of treatment regimen $(Z_1^{(b)}=z_1^{(b)},Z_2^{(b)}=z_2^{(b)})$ is denoted as $P^*_{z_2^{(b)}}=P_{z_2^{(b)}}P^*_{z_1^{(b)}}$.
	\item Estimate the model 
	\[Y_{z_1^{(b)},z_2^{(b)}}^{(b)}|P^*_{z_2^{(b)}},Z_1^{(b)}=z_1^{(b)},Z_2^{(b)}=z_2^{(b)},Y_{z_1^{(b)}}^{(b)},X_{z_1^{(b)}}^{(b)},X_{z_1^{(b)},z_2^{(b)}}^{(b)},W_1^{(b)},W_2^{(b)},V^{(b)}
	\] 
	again restricting to subjects that are observed and predicted to survive under the treatment regimes of interest at $t=2$. Use the respective models to impute the counterfactual of $Y_{z_1^{(b)},z_2^{(b)}}^{(b)}$.
	\item Using a similar procedure for steps 5-8 with the restriction determined by $S_{z_1^{(b)},\ldots,z_{t-1}^{(b)}}^{(b)}=S_{z_1^{'(b)},\ldots,z_{t-1}^{'(b)}}^{(b)}=1$ for time $t$ where at least one $z_t^{(b)}\neq z_t^{'(b)}$, extend the estimation until the desired time point $t=T$.
	\item Repeat Steps 1-9 to obtain $B$ bootstrap values for 
	\[
		\hat{\Delta}_{z_1^{(b)},\ldots,z_{t-1}^{(b)},z_1^{'(b)},\ldots,z_{t-1}^{'(b)}}^{(b)}=E[Y_{z_1^{(b)},\ldots,z_{t-1}^{(b)}}^{(b)}-Y_{z_1^{'(b)},\ldots,z_{t-1}^{'(b)}}^{(b)}|S_{z_1^{(b)},\ldots,z_{t-1}^{(b)}}^{(b)}=S_{z_1^{'(b)},\ldots,z_{t-1}^{'(b)}}^{(b)}=1].
	\]
	with associated pooled variance $W_{z_1^{(b)},\ldots,z_{t-1}^{(b)},z_1^{'(b)},\ldots,z_{t-1}^{'(b)}}^{(b)}$.
	\item The estimate of 
	\[
		\Delta_{Z_1,\ldots,Z_t,Z'_1,\ldots,Z'_t}=E[Y_{Z_1,\ldots,Z_t}-Y_{Z'_1,\ldots,Z'_t}|S_{Z_1,\ldots,Z_{t-1}}=S_{Z'_1,\ldots,Z'_{t-1}}=1]
	\] 
	is then 
	\[
		\bar{\Delta}_{z_1,\ldots,z_t,z'_1,\ldots,z'_t,B}=\sum_{b=1}^B(\hat{\Delta}_{z_1^{(b)},\ldots,z_{t-1}^{(b)},z_1^{'(b)},\ldots,z_{t-1}^{'(b)}}^{(b)})/B,
	\] 
	and the estimate of the variance of $\bar{\Delta}_{z_1,\ldots,z_t,z'_1,\ldots,z'_t,B}$ is 
	\[
		T_B=\bar{W}_{z_1,\ldots,z_t,z'_1,\ldots,z'_t,B}+(1+1/B)D_{z_1,\ldots,z_t,z'_1,\ldots,z'_t,B},
	\] 
	where 
	\[
		\bar{W}_{z_1,\ldots,z_t,z'_1,\ldots,z'_t,B}=\sum_{b=1}^B(W_{z_1^{(b)},\ldots,z_{t-1}^{(b)},z_1^{'(b)},\ldots,z_{t-1}^{'(b)}}^{(b)})/B 
	\]
	and
	\[
		D_{z_1,\ldots,z_t,z'_1,\ldots,z'_t,B}=\sum_{b=1}^B\frac{(\hat{\Delta}_{z_1^{(b)},\ldots,z_{t-1}^{(b)},z_1^{'(b)},\ldots,z_{t-1}^{'(b)}}^{(b)}-\bar{\Delta}_{z_1,\ldots,z_t,z'_1,\ldots,z'_t,B})^2}{B-1}. 
	\]
	The estimate $\Delta_{Z_1,\ldots,Z_t,Z'_1,\ldots,Z'_t}$ follows a $t$ distribution with degree of freedom $\nu$, 
	\[
		\frac{\Delta_{Z_1,\ldots,Z_t,Z'_1,\ldots,Z'_t}-\bar{\Delta}_{z_1,\ldots,z_t,z'_1,\ldots,z'_t,B}}{\sqrt{T_B}}\sim t_{\nu},
	\] 
	where $\nu=(B-1)(1+\frac{\bar{W}_{z_1,\ldots,z_t,z'_1,\ldots,z'_t,B}}{D_{z_1,\ldots,z_t,z'_1,\ldots,z'_t,B}(B+1)})^2$.
\end{enumerate}

\textit{Remark}. The idea of including the BART estimated propensity score within BART as a predictor in Steps 4 and 8 is not new. \cite{hahn} showed that including a BART estimated propensity score as a predictor within BART improved the estimation of heterogenous treatment effects for observational studies. \cite{tan} also reported that the inclusion of the BART estimated propensity score as a predictor within BART to impute missing data, under the missing at random assumption, worked well in situations where the non-linear main and interaction effects are complex for the mean and propensity model. For situations with simpler non-linear effects like a quadratic relationship, using BART to estimate the propensity score and imputing the missing values using penalized splines of propensity prediction \citep[][PENCOMP version for missing data]{zhang_little} worked better. Using PENCOMP with a BART estimated propensity score for Steps 4 and 8 would be an interesting alternative. However, our aim of Steps 4 and 8 was to ease the implementation burden on the researcher. Hence, we suggest the use of PENCOMP with a BART estimated propensity score for Steps 4 and 8 only if the researcher is certain that the non-linear effect has a simple form for example, a quadratic or cubic relationship.

\section{Simulation}
We conducted a simulation study to determine how well our proposed method would perform compared to the na\"{i}ve method and MSM in three scenarios: 1) where there is low association between treatment allocation and confounder as well as treatment and survival status; 2) where there is a strong association between treatment and confounder as well as treatment and survival status; and finally 3) where there is a strong association between treatment and confounder, treatment and survival status, and an interaction between treatment, confounder, and survival status. We expect all three methods to perform well in the first scenario because there is little to no confounding. For the second scenario, we expect MSM and our proposed method to perform well because there is no difference in the treatment effect between the principal strata,  and other stratification groups. The na\"{i}ve method should not perform well due to the strong association between treatment and confounder as well as treatment and survival status. Finally, for scenario three, we expect only our proposed method to perform well because an association between the treatment effect and principal strata, $S_{Z_1,\ldots,Z_{t-1}}=S_{Z'_1,\ldots,Z'_{t-1}}=1$, is induced by the stronger interaction effect between treatment, confounder, and survival status. We fit standard linear and logistic regression models rather than BART and PENCOMP with BART since our focus is not on model misspecification but rather, the effect of confounding by indication and censoring by death.

\subsection{Setup}
To set up our simulation study, we set the size of our target population as 1 million. We then generate a single baseline variable $V$ from a normal distribution. We set $T=3$ and model our treatment allocation, $Z_1$, as 
\begin{equation}
	\label{chap3:ztime1}
	logit[P(Z_1=1|V)]=\gamma_0+\gamma_1V.
\end{equation}
For the potential outcome at $t=1$, $Y_{Z_1}$, we model it as
\begin{equation}
	\label{chap3:ytime1}
	Y_{Z_1}=\beta_0+\beta_ZI\{Z_1=1\}+\beta_VV+\beta_{VZ}VI\{Z_1=1\}+e,
\end{equation}
where $e\sim N(0,1)$. 

We model the potential survival status at $t=2$, $S_{Z_1}$ as
\begin{align}
	\label{chap3:stime2}
	logit(P[S_{Z_1}=1|V,Y_{Z_1}])&=\alpha_0+\alpha_{Y_1}Y_1I\{Z_1=1\}+\alpha_{Y_0}Y_0[1-I\{Z_1=1\}] \nonumber\\
	&\quad+\alpha_ZI\{Z_1=1\}+\alpha_VV+\alpha_{VZ}VI\{Z_1=1\}.
\end{align}
Monotonicity is imposed by setting $S_0=1$ if $S_1=1$. Because a negative wealth shock is an absorbing state, if $Z_1=1$, then $Z_2=1$. So when $Z_1=0$, we have 
\begin{equation}
	\label{chap3:ztime2}
	logit(P[Z_2=1|V,Y_0])=\gamma_0+\gamma_{Y_0,2}Y_0+\gamma_2V.
\end{equation}
We model the potential outcome at $t=2$, $Y_{Z_1,Z_2}$ as
\begin{align}
	\label{chap3:ytime2}
	Y_{Z_1,Z_2}&=\beta_0+\beta_{Z_{01}}I\{Z_1=0,Z_2=1\}+\beta_{Z_{11}}I\{Z_1=1,Z_2=1\}\nonumber\\
	&\quad+\beta_{Y_0Z_{00}}Y_0I\{Z_1=0,Z_2=0\}+\beta_{Y_0Z_{01}}Y_0I\{Z_1=0,Z_2=1\}\nonumber\\
	&\quad+\beta_{Y_1Z_{11}}Y_1I\{Z_1=1,Z_2=1\}+\beta_VV+\beta_{VZ_{01}}VI\{Z_1=0,Z_2=1\}\nonumber\\
	&\quad+\beta_{VZ_{11}}VI\{Z_1=1,Z_2=1\}+e,
\end{align}
where $e\sim N(0,1)$.

For the potential survival status at $t=3$, $S_{Z_1,Z_2}$, if $S_{Z_1}=0$, then $S_{Z_1,Z_2}=0$. When $S_{Z_1}=1$, we have
\begin{align}
	\label{chap3:stime3}
	logit(P[S_{Z_1,Z_2}=1|X,Y_{Z_1,Z_2},S_{Z_1}=1])&=\alpha_0+\alpha_{Z_{01}}I\{Z_1=0,Z_2=1\}\nonumber\\
	&\quad+\alpha_{Z_{11}}I\{Z_1=1,Z_2=1\}\nonumber\\
	&\quad+\alpha_{Y_{00}Z_{00}}Y_{00}I\{Z_1=0,Z_2=0\}\nonumber\\
	&\quad+\alpha_{Y_{01}Z_{01}}Y_{01}I\{Z_1=0,Z_2=1\}\nonumber\\
	&\quad+\alpha_{Y_{11}Z_{11}}Y_{11}I\{Z_1=1,Z_2=1\}\nonumber\\
	&\quad+\alpha_VV+\alpha_{VZ_{01}}VI\{Z_1=0,Z_2=1\}\nonumber\\
	&\quad+\alpha_{VZ_{11}}VI\{Z_1=1,Z_2=1\}.
\end{align}
Again, we impose monotonicity by setting $S_{00}=S_{01}=1$ if $S_{11}=1$ and $S_{00}=1$ if $S_{01}=1$. For the treatment allocation at $t=3$, $Z_3$, if $Z_1=Z_2=0$, we have
\begin{equation}
	\label{chap3:ztime3}
	logit(P[Z_3=1|X,Y_{00}])=\gamma_0+\gamma_{Y_{00}}Y_{00}+\gamma_{Y_{0,3}}Y_0+\gamma_3V.
\end{equation}
For the potential outcome at $t=3$, $Y_{Z_1,Z_2,Z_3}$, we have
\begin{align}
	\label{chap3:ytime3}
	Y_{Z_1,Z_2,Z_3}&=\beta_0+\beta_{Z_{001}}I\{Z_1=0,Z_2=0,Z_3=1\}+\beta_{Z_{011}}I\{Z_1=0,Z_2=1,Z_3=1\}\nonumber\\
	&\quad+\beta_{Z_{111}}I\{Z_1=1,Z_2=1,Z_3=1\}+\beta_{Y_{00}Z_{000}}Y_{00}I\{Z_1=0,Z_2=0,Z_3=0\}\nonumber\\
	&\quad+\beta_{Y_{00}Z_{001}}Y_{00}I\{Z_1=0,Z_2=0,Z_3=1\}\nonumber\\
	&\quad+\beta_{Y_{01}Z_{011}}Y_{01}I\{Z_1=0,Z_2=1,Z_3=1\}\nonumber\\
	&\quad+\beta_{Y_{11}Z_{111}}Y_{11}I\{Z_1=1,Z_2=1,Z_3=1\}+\beta_{Y_0Z_0}Y_0I\{Z_1=0\}\nonumber\\
	&\quad+\beta_{Y_1Z_1}Y_1I\{Z_1=1\}+\beta_VV+\beta_{VZ_{001}}VI\{Z_1=0,Z_2=0,Z_3=1\}\nonumber\\
	&\quad+\beta_{VZ_{011}}VI\{Z_1=0,Z_2=1,Z_3=1\}\nonumber\\
	&\quad+\beta_{VZ_{111}}VI\{Z_1=1,Z_2=1,Z_3=1\}+e.
\end{align}

Table \ref{chap3:sim_para1} shows the parameters we used to achieve the three different simulation scenarios. Scenario 1 is achieved by setting $\gamma_1$, $\alpha_Z$, $\gamma_2$, $\gamma_{Y_0,2}$, $\alpha_{Z_{01}}$, $\alpha_{Z_{11}}$, $\gamma_3$, $\gamma_{Y_0,3}$, and $\gamma_{Y_{00}}$ to be about 10 times smaller than the values in Scenarios 2 and 3. The rest of the differences between Scenario 1 versus 2 and 3 were to ensure the resulting simulated population would have enough deaths and subjects in the various different treatment regimes for the assumptions used by MSM and our proposed method to be valid. The difference between Scenario 2 versus 3 lie in $\beta_{VZ}$, $\alpha_{Y_1}$, $\alpha_{Y_0}$, $\beta_{Y_0Z_{00}}$, $\beta_{Y_0Z_{01}}$, $\beta_{Y_1Z_{11}}$, $\alpha_{Y_0Z_{00}}$, $\alpha_{Y_0Z_{01}}$, $\alpha_{Y_1Z_{11}}$, $\beta_{Y_{00}Z_{000}}$, $\beta_{Y_{00}Z_{001}}$, $\beta_{Y_{01}Z_{011}}$, and $\beta_{Y_{11}Z_{111}}$ where the values for Scenario 2 is about 10 times smaller compared to Scenario 3.

To calculate the true parameters, we used the generated population data (size 1 million), and then took:
\begin{enumerate}
	\item $\Delta_{1,0}=\bar{Y}_1-\bar{Y}_0$;
	\item $\Delta_{01,00}=\bar{Y}_{01}-\bar{Y}_{00}$ given $S_0=1$;
	\item $\Delta_{11,00}=\bar{Y}_{11}-\bar{Y}_{00}$ given $S_0=S_1=1$;
	\item $\Delta_{11,01}=\bar{Y}_{11}-\bar{Y}_{01}$ given $S_0=S_1=1$;
	\item $\Delta_{001,000}=\bar{Y}_{001}-\bar{Y}_{000}$ given $S_{00}=1$;
	\item $\Delta_{011,000}=\bar{Y}_{011}-\bar{Y}_{000}$ given $S_{00}=S_{01}=1$;
	\item $\Delta_{111,000}=\bar{Y}_{111}-\bar{Y}_{000}$ given $S_{00}=S_{11}=1$;
	\item $\Delta_{011,001}=\bar{Y}_{011}-\bar{Y}_{001}$ given $S_{00}=S_{01}=1$;
	\item $\Delta_{111,001}=\bar{Y}_{111}-\bar{Y}_{001}$ given $S_{00}=S_{11}=1$; and 
	\item $\Delta_{111,011}=\bar{Y}_{111}-\bar{Y}_{011}$ given $S_{01}=S_{11}=1$.
\end{enumerate}

We measured performance using the empirical bias, root mean squared error (RMSE), 95\% coverage, and the average 95\% Confidence Interval (CI) length (AIL). 1000 simulations were used to estimate these quantities. Under each simulation, a simple random sample of 4,000 or 8,000 subjects was drawn from the target population data. All methods were then implemented on the sampled data to obtain the effect estimates. For MSM and our proposed method, the models were specified using Equations \ref{chap3:ztime1} to \ref{chap3:ytime3} respectively. For our proposed method, because our focus is not on model misspecification but rather, confounding by indication and censoring by death, we chose to implement a simpler version of our method by  skipping Steps 3 and 7 of our algorithm and using $Y_{z_1^{(b)}}^{(b)}|Z_1^{(b)}=z_1^{(b)},X_{z_1^{(b)}}^{(b)},W_1^{(b)},V^{(b)}$ and $Y_{z_1^{(b)},z_2^{(b)}}^{(b)}|Z_1^{(b)}=z_1^{(b)},Z_2^{(b)}=z_2^{(b)},Y_{z_1^{(b)}}^{(b)},X_{z_1^{(b)}}^{(b)},X_{z_1^{(b)},z_2^{(b)}}^{(b)},W_1^{(b)},W_2^{(b)},V^{(b)}$ for Steps 4 and 8 respectively. We also simplified the prediction of the potential outcomes and survival status by using linear and logistic regression instead of BART.

\subsection{Results}
Table \ref{chap3:sim_res1} shows the simulation results for sample size of 4,000. As expected, under Scenario 1, all three methods were relatively unbiased with all three methods achieving similar RMSE. MSM and our proposed method reported slightly greater than nominal coverage due to the wider AIL. Under Scenario 2, the absolute bias of the na\"{i}ve method was always larger than MSM and our proposed method. RMSE was larger as well in comparison and coverage was often far below the nominal 95\% value. For this scenario MSM produced the less conservative coverage while our proposed method suggested better bias performance and reduced RMSE. Finally, under Scenario 3, the na\"{i}ve method was clearly biased with poor RMSE and coverage. MSM performed slightly better compared to the na\"{i}ve method but absolute bias clearly increased compared to Scenario 2. Coverage for some treatment effects were poor as well. Our proposed method remained unbiased, produced a lower RMSE compared to the other two methods, and reached nominal coverage under Scenario 3. All methods behaved as expected under these three scenarios.

Table \ref{chap3:sim_res2} shows the results with the sample size increased to 8,000, approximately the sample size in our application. The simulation results for all three methods under Scenario 1 remained relatively similar. Under Scenario 2, an increase in sample size did not affect the absolute bias of all three methods but, the coverage of the na\"{i}ve method was clearly affected with huge decreases in the coverage for all parameters. Coverage for MSM and our proposed method remained fairly similar. Finally, under Scenario 3, we observe once again that the amount of bias for the three methods remained the same but, coverage for the na\"{i}ve method and MSM decreased for most of the treatment effects when the sample size increased to 8,000. Coverage for our proposed method remained relatively similar to the results observed for the sample size of 4,000. In summary, bias for the three methods was rather stable when the sample size changed. However, if the method is poor in the estimation of the particular treatment effect, increasing the sample size can cause large decreases in coverage.

\section{Determining the effect of a negative wealth shock on cognitive score for Health and Retirement Study subjects}
\subsection{Health and Retirement Study}
To investigate the association between negative wealth shock and cognitive ability in late middle aged US adults, we used data from the Health and Retirement Study (HRS). HRS is a longitudinal study of US adults, enrolled at age 50 and older. These individuals have been surveyed biennially since 1992 with detailed modules on financial status and health \citep{sonnega}.

We use HRS data collected from 1996 to 2002 for our analysis. Subjects were obtained from the original HRS cohort, born in the years 1931-1941. Although data collection began in 1992, consistent collection of a subject's cognitive ability only began in 1996. Hence, we excluded the data collected before 1996 and treated the variables collected in 1996 as the baseline for our analysis. We excluded subjects who did not have longitudinal measurements for net worth because we were unable to distinguish whether they have already experienced a negative wealth shock. Subjects with zero or negative net worth at baseline were excluded since we did not know if these subjects have lifelong asset poverty or experienced a negative wealth shock prior to study entry. We also removed subjects who experienced a negative wealth shock and death between 1992 to 1996. These subjects were removed because they were no longer at risk for a negative wealth shock or death. There were 9,750 participants in the original HRS cohort, and of these, 7,106 participants (72.9\%) were eligible for this analysis. These participants consists of a representative sample of the 1996 US population aged 55 to 65 who had not experienced a negative wealth shock in the previous five years.

\subsubsection{Determining negative wealth shock}
To determine whether a subject experienced a negative wealth shock from the previous follow-up period to the current follow-up period, we first obtained data from the module assessing net worth administered at every wave of HRS. Measured assets include housing value, net value of businesses, individual retirement accounts, checking/savings accounts, certificates of deposits and savings bonds, investment holdings, net value of vehicles, and the value of any other substantial assets. From this asset total, debts were subtracted, including home mortgages, other home equity loans, and unsecured debt values, like credit card balances, student loans, and medical debts. Missing values for wealth were imputed at the level of each asset or debt, using an unfolding bracket imputation method \citep{juster}. Wealth data were not imputed for those who do not participate in a given wave. Negative wealth shock was measured and then dichotomized (yes or no) for each time point. Loss of 75\% or more of total wealth between two consecutive waves was used as the cut-point for negative wealth shock \citep{pool2}. Subjects were considered at risk for negative wealth shock until they have experienced a negative wealth shock or reached age 65.  
 
\subsubsection{Cognitive ability}
The cognitive ability of a subject is assessed in HRS using the Telephone Interview for Cognitive Status (TICS). Unfortunately, the full HRS cognitive battery is not available for participants under 65. Hence, we used an abbreviated measure that included questions about episodic memory (Immediate Word recall [10 points] and Delayed Word recall [10 points]) and mental status (Serial 7's [5 points], backwards counting from 20 [2 points]) \citep{crimmins}. All responses were combined to create a composite score ranging from 0 to 27, with a higher score indicating higher cognitive ability. Some of these measures may be imputed implying that the cognitive summary score may include one or more imputed scores \citep{fisher}. We treated this measure as continuous and normally distributed.

\subsubsection{Descriptive statistics at baseline}
Tables \ref{chap3:des_stat1} to \ref{chap3:des_stat2} show the descriptive statistics of the subjects at baseline by whether or not they experienced a negative wealth shock over the next six years regardless of survival status. At baseline, aside from whether the subject eventually survived until 2002 and health conditions like whether the subject ever had heart problems, high blood pressure, and stroke, all the other variables in Tables \ref{chap3:des_stat1} to \ref{chap3:des_stat2} were significantly associated with experiencing a negative wealth shock. A typical subject who would eventually experience a wealth shock would have a lower cognitive score at baseline; slightly higher BMI; lower opinion about his or her health; lower word recall score; likely still smoking; not insured; have depression; slightly lower income; either working, unemployed, or disabled; divorced or never married; lower wealth rank; have diabetes and/or psychological problems; younger; lesser years of education; and likely non-White.

Table \ref{chap3:cog_score_ts} shows the change in unadjusted mean cognitive score between consecutive waves for subjects who did not receive a wealth shock versus those who ever received a negative wealth shock. Follow-up surveys occurred at years 2, 4, and 6. We can see that for a subject who ever got shocked, the largest observed decline in cognitive score occurs from Baseline to Wave 1. Subsequently, the decline in cognitive score is no longer as large between waves. Similarly, the bulk of our subjects were shocked at Wave 1 (second year of follow up). In later waves, the proportion of new subjects who received a negative wealth shock decreases.

\subsection{Analysis}
We were interested in how a negative wealth shock would affect the cognitive ability of late middle aged adults in the HRS during the six years of follow-up as well as how the duration of a negative wealth shock affects cognitive ability accounting for missingness in the cognitive outcome as well as censoring by death. We employed four different methods to estimate this effect and make inference. The four methods were the na\"{i}ve method, where all subjects who died under their observed negative wealth shock status were removed from analysis; baseline adjusted method, where similar to the na\"{i}ve method, all subjects who died were removed from analysis but the mean cognitive score was adjusted using a model that included all baseline covariates; MSM, where negative wealth shock allocation, missingness, and censoring by death were accounted for by inverse probability weighting; and our proposed method including the PENCOMP modification described in Subsection \ref{chap3:prop_meth}. We assumed that depression was the time-varying covariate that depends on the negative wealth shock status ($X_{Z_1,\ldots,Z_t}$ in Section \ref{chap3:review}) and the rest of the time-varying covariates are: self-reported health status, whether subject was insured, labor force status of subject, income, level of alcohol consumption, current smoking status, and number of health conditions ($W_t$ in Section \ref{chap3:method}). We also assumed that the cognitive score is missing at random given the baseline variables presented in Tables \ref{chap3:des_stat1} to \ref{chap3:des_stat2}, past negative wealth shock status, time-varying covariates, and cognitive score. For MSM, we accounted for this missingness by modeling the propensity of response while for our proposed method, we imputed the missing cognitive score by using the modified version of PENCOMP discussed in Subsection \ref{chap3:prop_meth}. All our models (baseline adjusted, MSM, and our proposed method) were specified using BART. For the na\"{i}ve, baseline adjusted, and MSM method, we employed 1,000 bootstrap samples to calculate the mean and the 95\% Confidence Interval (CI). The 95\% CI was determined by taking the $2.5$ and $97.5$ percentile. For our proposed method, we estimated the effect and accounted for our uncertainty using our algorithm described in Subsection \ref{chap3:prop_meth}.
   
\subsection{Results}
Table \ref{chap3:anal_res} shows the adjusted effect estimate of a negative wealth shock on cognitive score depending on the duration of the shock for late middle aged adults in the original HRS cohort from 1996 to 2002. In general, the na\"{i}ve and baseline adjusted method suggests that experiencing a negative wealth shock has a much larger negative effect on the cognitive score of subjects in our sample compared to the adjusted estimates reported by MSM and our proposed method. The na\"{i}ve and baseline adjusted method produced very similar results suggesting low association between cognitive score and the baseline covariates. The effect for subjects who experienced a negative wealth shock within the first 2 years of follow up versus no shock (6 years vs. no shock), subjects who experienced a negative wealth shock within the first 2 years of follow up versus subjects who experienced a negative wealth shock between the second and fourth year of follow up (6 years vs. 2 years), and subjects who experienced a negative wealth shock within the first 2 years of follow up versus subjects who experienced a negative wealth shock between the fourth and sixth year of follow up (6 years vs. no shock), were significantly larger than 0 under the na\"{i}ve and baseline adjusted method. For MSM and our proposed method all effects were reported to be not significant.
 
\section{Discussion}
In this paper, we were interested in how a negative wealth shock affects the cognitive ability of late middle aged Americans participating in the HRS from 1996 to 2002. The main difficulty we faced was the presence of death in some subjects causing their cognitive score to be censored. Under situations where we believe death does not depend on the cognitive ability or whether a subject received a negative wealth shock, removing subjects who have died from our analysis would yield an unbiased estimate of the effect of negative wealth shock on cognitive ability as our simulation results suggest. Unfortunately, it is very possible that subjects with lower cognitive ability and/or have experienced a negative wealth shock would have a higher risk of death. In this situation, accounting for the censoring by death would be needed. This is because without randomization, there is a high likelihood that the proportion of deaths between subjects who did not receive a negative wealth shock versus those who received a wealth shock, would be imbalanced. In addition, subjects who die are more likely to have a lower cognition score. As a result, if we remove the subjects who died from our analysis, the effect of the negative wealth shock on cognitive ability that we measure would be confounded by death. Although MSM is commonly employed to weight the subjects who survived, this approach is arguably not sensible and would likely produce biased estimates when the effect depends on the principal strata as well as when adjustments on the weights have to be employed in order to stabilize the MSM estimate. To overcome these issues, we propose a new method to estimate the effect by imputing the counterfactual survival status of each subject in order to compare outcomes among individuals who would survive only under both sets of treatments being considered. Our method remained unbiased for all the simulation scenarios we tried and produced reasonable coverage. When applied to the HRS dataset, our method suggested that the effect of a negative wealth shock on the cognitive ability is close to null whereas the na\"{i}ve method and MSM suggested an estimate with a slightly larger effect.

One shortcoming of our approach is our failure to incorporate the HRS sample design, in particular the sampling weights, in our inference. Given that a key use of weights in regression-type analysis is to reduce the effect of model misspecification \citep{korn}, we hope that our use of BART will minimize the degree of model misspecification. We leave the incorporation of such features in a general approach to future work. Another aspect of our method which could be improved is to allow our method to be applicable to studies where the follow-up time is not fixed. In such a situation, Cox based survival models would have to be employed and time would have to be included as a covariate in the survival and outcome models. The difficulty in this extension would be how to develop a systematic way, applicable to all subjects, to determine the relation in time between the allocation of the treatment, measuring the outcome, and death. 

%\section*{Supplementary materials}
%The codes and dataset to obtain Tables \ref{chap3:sim_res1} to \ref{chap3:cog_score_ts} are contained in the zip file ``Supplementary materials.zip'' available online.

\bibliography{references}
\bibliographystyle{biom}

\begin{landscape}
\begin{table}
	\caption{Sample example of a censoring by death dataset until $t=3$ where $Z_t=1$ indicates a subject having experienced a negative wealth shock and $Z_t=0$ indicates a subject have not experienced any negative wealth shock till time $t$  \label{chap3:eg2_1}}
	\centering
	\begin{tabular}{|c|c|c|c|c|c|c|c|c|c|c|c|c|c|c|c|c|c|c|c|}
	\hline
		& $V$ & $Z_1$ & $Y_1$ & $Y_0$ & $S_1$ & $S_0$ & $Z_2$ & $Y_{00}$ & $Y_{01}$  & $Y_{11}$ & $S_{00}$ & $S_{01}$  & $S_{11}$ & $Z_3$ & $Y_{000}$ & $Y_{001}$  & $Y_{011}$ & $Y_{111}$\\
		\hline
		Subject 1 & x & 1 & x & ? & 1 & 1 & 1 & ? & ? & x & 1 & 1 & 1 & 1 & ? & ? & ? & x\\
		\hline
		Subject 2 & x & 0 & ? & x & 1 & 1 & 1 & ? & x & ? & 1 & 1 & 1 & 1 & ? & ? & x & ?\\
		\hline
		Subject 3 & x & 1 & x & ? & 1 & 1 & 1 & ? & ? & x & 1 & 1 & 0 & NA & ? & ? & ? & NA\\
		\hline
		Subject 4 & x & 0 & ? & x & 1 & 1 & 1 & ? & x & ? & 1 & 1 & 0 & 1 & ? & ? & x & NA\\
		\hline
		Subject 5 & x & 0 & ? & x & 1 & 1 & 0 & x & ? & ? & 1 & 0 & 1 & 0 & x & ? & NA & ?\\
		\hline
		Subject 6 & x & 0 & ? & x & 1 & 1 & 0 & x & ? & ? & 0 & 1 & 1 & NA & NA & NA & ? & ?\\
		\hline
		Subject 7 & x & 0 & ? & x & 1 & 1 & 0 & x & ? & ? & 0 & 1 & 1 & NA & NA & NA & ? & ?\\
		\hline
		Subject 8 & x & 0 & ? & x & 1 & 1 & 0 & x & ? & ? & 1 & 0 & 0 & 0 & x & ? & NA & NA\\
		\hline		
		Subject 9 & x & 1 & x & ? & 0 & 1 & NA & ? & ? & NA & 1 & 1 & 0 & NA & ? & ? & ? & NA \\
		\hline
		Subject 10 & x & 1 & x & ? & 0 & 1 & NA & ? & ? & NA & 0 & 1 & 0 & NA & NA & NA & ? & NA \\
		\hline
		Subject 11 & x & 0 & ? & x & 0 & 1 & 1 & ? & x & NA & 0 & 1 & 0 & 1 & NA & NA & x & NA\\
		\hline
		Subject 12 & x & 0 & ? & x & 0 & 1 & 0 & x & ? & NA & 0 & 1 & 0 & NA & NA & NA & ? & NA\\
		\hline	
		Subject 13 & x & 1 & x & ? & 1 & 0 & 1 & NA & NA & x & 0 & 0 & 1 & 1 & NA & NA & NA & x\\
		\hline
		Subject 14 & x & 0 & ? & x & 1 & 0 & NA & NA & NA & ? & 0 & 0 & 1 & NA & NA & NA & NA & ?\\
		\hline
	\end{tabular}
\end{table} 
\end{landscape}

\begin{table}[H]
	\caption{Table of parameters for simulation \label{chap3:sim_para1}}
	\tiny
	\centering
	\begin{tabular}{|c|c|c|c|}
	\hline
	& Scenario 1 & Scenario 2 & Scenario 3 \\
	\hline
	$V$ & $N(0,2^2)$ & $N(17,2^2)$ & $N(17,2^2)$ \\
	$\gamma_0$ & 0 & 2 & 2 \\
	$\gamma_1$ & -0.02 & -0.2 & -0.2 \\
	$\beta_0$ & 0 & 5.3 & 5.3 \\
	$\beta_Z$ & -1.5 & -1.5 & -1.5 \\
	$\beta_V$ & 0.015 & 0.15 & 0.2 \\
	$\beta_{VZ}$ & -0.005 & -0.11 & -0.05\\
	$\alpha_0$ & 0 & 1 & 0 \\
	$\alpha_{Y_1}$ & 0.005 & 0.00625 & 0.0625 \\
	$\alpha_{Y_0}$ & 0.01 & 0.0125 & 0.125 \\
	$\alpha_Z$ & -0.01 & -0.2 & -0.2 \\
	$\alpha_V$ & 0.002 & 0.02 & 0.02 \\
	$\alpha_{VZ}$ & -0.002 & -0.02 & -0.02\\
	$\gamma_2$ & -0.002 & -0.02 & -0.02\\
	$\gamma_{Y_0,2}$ & -0.02 & -0.2 & -0.2\\
	$\beta_{Z_{01}}$ & -1.5 & -1.5 & -1.5\\
	$\beta_{Z_{11}}$ & -1 & -1 & -1\\
	$\beta_{Y_0Z_{00}}$ & 0.015 & 0.02 & 0.3\\
	$\beta_{Y_0Z_{01}}$ & 0.01 & 0.015 & 0.2\\
	$\beta_{Y_1Z_{11}}$ & 0.005 & 0.01 & 0.1\\
	$\beta_{VZ_{01}}$ & -0.00011 & -0.011 & -0.011\\
	$\beta_{VZ_{11}}$ & -0.00005 & -0.005 & -0.005\\
	$\alpha_{Z_{01}}$ & -0.01 & -0.2 & -0.2 \\
	$\alpha_{Z_{11}}$ & -0.015 & -0.1 & -0.1 \\
	$\alpha_{Y_0Z_{00}}$ & 0.01 & 0.0125 & 0.125\\
	$\alpha_{Y_0Z_{01}}$ & 0.005 & 0.00625 & 0.0625\\
	$\alpha_{Y_1Z_{11}}$ & 0.0025 & 0.003125 & 0.03125\\
	$\alpha_{VZ_{01}}$ & -0.0001 & -0.02 & -0.02\\
	$\alpha_{VZ_{11}}$ & -0.0005 & -0.05 & -0.05\\
	$\gamma_3$ & -0.0002 & -0.002 & -0.002\\
	$\gamma_{Y_0,3}$ & -0.002 & -0.02 & -0.02\\
	$\gamma_{Y_{00}}$ & -0.02 & -0.2 & -0.2\\
	$\beta_{Z_{001}}$ & -1.5 & -1.5 & -1.5\\
	$\beta_{Z_{011}}$ & -1 & -1 & -1\\
	$\beta_{Z_{111}}$ & -0.5 & -0.5 & -0.5\\
	$\beta_{Y_{00}Z_{000}}$ & 0.015 & 0.02 & 0.3\\
	$\beta_{Y_{00}Z_{001}}$ & 0.01 & 0.015 & 0.2\\
	$\beta_{Y_{01}Z_{011}}$ & 0.005 & 0.01 & 0.1\\
	$\beta_{Y_{11}Z_{111}}$ & 0.0025 & 0.005 & 0.05\\
	$\beta_{Y_0Z_0}$ & 0.0008 & 0.08 & 0.08\\
	$\beta_{Y_1Z_1}$ & 0.0003 & 0.03 & 0.03\\
	$\beta_{VZ_{001}}$ & -0.00011 & -0.011 & -0.011\\
	$\beta_{VZ_{011}}$ & -0.00005 & -0.005 & -0.005\\
	$\beta_{VZ_{111}}$ & -0.00003 & -0.003 & -0.003\\
	\hline
	\end{tabular}
\end{table}

\begin{table}[H]
	\caption{Simulation results for sample size 4,000 \label{chap3:sim_res1}}
	\hspace{-1.75cm}
	\tiny
	\begin{tabular}{|lc|cccc|cccc|cccc|}
	\hline
	 \multicolumn{2}{|c|}{Scenario 1} & \multicolumn{4}{c|}{Na\"{i}ve} & \multicolumn{4}{c|}{MSM} & \multicolumn{4}{c|}{Proposed}\\
	Parameter & True value & Bias & RMSE & 95\% Coverage & AIL & Bias & RMSE & 95\% Coverage & AIL & Bias & RMSE & 95\% Coverage & AIL \\ 
	\hline
	$\Delta_{1,0}$ & -1.497 & -0.001 & 0.032 & 95.4 & 0.123 & -0.0002 & 0.032 & 95.1 & 0.123 & -0.0001 & 0.032 & 97.0 & 0.143 \\
	$\Delta_{01,00}$ & -1.499 & -0.003 & 0.050 & 95.3 & 0.202 & -0.003 & 0.050 & 95.3 & 0.202 & -0.003 & 0.050 & 95.7 & 0.214 \\
	$\Delta_{11,00}$ & -1.005 & -0.003 & 0.049 & 95.0 & 0.189 & -0.001 & 0.049 & 94.7 & 0.189 & -0.001 & 0.049 & 99.2 & 0.262 \\
	$\Delta_{11,01}$ & 0.493 & 0.002 & 0.048 & 94.4 & 0.189 & 0.003 & 0.048 & 94.5 & 0.189 & 0.002 & 0.049 & 99.1 & 0.262 \\
	$\Delta_{001,000}$ & -1.502 & 0.005 & 0.081 & 93.9 & 0.314 & 0.005 & 0.081 & 98.9 & 0.411 & 0.005 & 0.082 & 94.6 & 0.333 \\
	$\Delta_{011,000}$ & -1.008 & 0.004 & 0.074 & 94.8 & 0.284 & 0.004 & 0.074 & 99.0 & 0.370 & 0.004 & 0.075 & 97.8 & 0.350 \\
	$\Delta_{111,000}$ & -0.504 & 0.006 & 0.072 & 95.2 & 0.284 & 0.007 & 0.072 & 99.4 & 0.371 & 0.007 & 0.074 & 100.0 & 0.529 \\
	$\Delta_{011,001}$ & 0.495 & -0.001 & 0.071 & 95.0 & 0.284 & -0.0001 & 0.072 & 99.1 & 0.370 & -0.0009 & 0.072 & 97.8 & 0.348 \\
	$\Delta_{111,001}$ & 1.000 & -0.0001 & 0.072 & 95.6 & 0.284 & 0.001 & 0.072 & 99.0 & 0.371 & 0.001 & 0.074 & 99.9 & 0.528 \\
	$\Delta_{111,011}$ & 0.502 & 0.003 & 0.065 & 94.3 & 0.250 & 0.005 & 0.065 & 98.9 & 0.325 & 0.005 & 0.067 & 99.9 & 0.440 \\
	\hline
	 \multicolumn{2}{|c|}{Scenario 2} & \multicolumn{4}{c|}{Na\"{i}ve} & \multicolumn{4}{c|}{MSM} & \multicolumn{4}{c|}{Proposed}\\
	Parameter & True value & Bias & RMSE & 95\% Coverage & AIL & Bias & RMSE & 95\% Coverage & AIL & Bias & RMSE & 95\% Coverage & AIL \\ 
	\hline
	$\Delta_{1,0}$ & -3.367 & -0.047 & 0.061 & 78.5 & 0.154 & 0.002 & 0.041 & 93.8 & 0.160 & 0.002 & 0.041 & 96.1 & 0.177 \\
	$\Delta_{01,00}$ & -1.727 & -0.037 & 0.054 & 83.2 & 0.149 & -0.032 & 0.051 & 86.9 & 0.150 & -0.002 & 0.037 & 96.4 & 0.161 \\
	$\Delta_{11,00}$ & -1.199 & -0.136 & 0.146 & 24.5 & 0.202 & -0.020 & 0.057 & 92.5 & 0.204 & -0.004 & 0.053 & 96.5 & 0.229 \\
	$\Delta_{11,01}$ & 0.528 & -0.098 & 0.111 & 49.2 & 0.199 & 0.013 & 0.054 & 93.7 & 0.201 & -0.001 & 0.053 & 97.0 & 0.226 \\
	$\Delta_{001,000}$ & -1.727 & -0.029 & 0.062 & 91.9 & 0.220 & -0.023 & 0.060 & 94.8 & 0.240 & 0.001 & 0.053 & 96.1 & 0.227 \\
	$\Delta_{011,000}$ & -1.183 & -0.065 & 0.082 & 75.0 & 0.199 & -0.047 & 0.069 & 87.5 & 0.217 & 0.0004 & 0.048 & 97.8 & 0.220 \\
	$\Delta_{111,000}$ & -1.169 & -0.167 & 0.181 & 33.8 & 0.273 & -0.042 & 0.084 & 93.2 & 0.305 & -0.004 & 0.071 & 98.7 & 0.350 \\
	$\Delta_{011,001}$ & 0.544 & -0.036 & 0.059 & 88.0 & 0.185 & -0.024 & 0.053 & 94.4 & 0.202 & -0.002 & 0.045 & 96.7 & 0.206 \\
	$\Delta_{111,001}$ & 0.558 & -0.139 & 0.153 & 45.7 & 0.264 & -0.019 & 0.071 & 96.3 & 0.294 & -0.007 & 0.067 & 98.4 & 0.331 \\
	$\Delta_{111,011}$ & 0.013 & -0.101 & 0.119 & 62.9 & 0.246 & 0.007 & 0.065 & 96.1 & 0.276 & -0.002 & 0.063 & 98.1 & 0.299 \\
	\hline
	 \multicolumn{2}{|c|}{Scenario 3} & \multicolumn{4}{c|}{Na\"{i}ve} & \multicolumn{4}{c|}{MSM} & \multicolumn{4}{c|}{Proposed}\\
	Parameter & True value & Bias & RMSE & 95\% Coverage & AIL & Bias & RMSE & 95\% Coverage & AIL & Bias & RMSE & 95\% Coverage & AIL \\ 
	\hline
	$\Delta_{1,0}$ & -2.347 & -0.123 & 0.130 & 14.5 & 0.160 & 0.002 & 0.042 & 94.0 & 0.160 & 0.002 & 0.042 & 95.9 & 0.177 \\
	$\Delta_{01,00}$ & -2.559 & -0.114 & 0.122 & 23.9 & 0.165 & -0.060 & 0.074 & 70.2 & 0.164 & -0.001 & 0.038 & 96.4 & 0.163 \\
	$\Delta_{11,00}$ & -3.062 & -0.231 & 0.239 & 2.8 & 0.232 & -0.033 & 0.068 & 89.9 & 0.226 & -0.004 & 0.058 & 97.0 & 0.260 \\
	$\Delta_{11,01}$ & -0.502 & -0.118 & 0.132 & 48.9 & 0.233 & 0.026 & 0.065 & 92.2 & 0.227 & -0.003 & 0.059 & 96.7 & 0.260 \\
	$\Delta_{001,000}$ & -2.820 & -0.125 & 0.139 & 47.9 & 0.242 & -0.062 & 0.087 & 88.7 & 0.273 & -0.0004 & 0.054 & 95.9 & 0.224 \\
	$\Delta_{011,000}$ & -3.605 & -0.143 & 0.152 & 19.6 & 0.198 & -0.087 & 0.101 & 69.2 & 0.225 & -0.006 & 0.045 & 96.4 & 0.202 \\
	$\Delta_{111,000}$ & -4.032 & -0.290 & 0.301 & 5.2 & 0.319 & -0.082 & 0.117 & 89.4 & 0.376 & -0.009 & 0.080 & 98.6 & 0.400 \\
	$\Delta_{011,001}$ & -0.785 & -0.019 & 0.060 & 93.3 & 0.225 & -0.026 & 0.063 & 95.0 & 0.256 & -0.006 & 0.052 & 96.7 & 0.226 \\
	$\Delta_{111,001}$ & -1.217 & -0.160 & 0.181 & 54.4 & 0.336 & -0.015 & 0.087 & 97.2 & 0.396 & -0.009 & 0.083 & 99.3 & 0.442 \\
	$\Delta_{111,011}$ & -0.432 & -0.141 & 0.160 & 54.9 & 0.306 & 0.011 & 0.080 & 97.4 & 0.363 & -0.006 & 0.075 & 98.7 & 0.373 \\
	\hline
	\end{tabular}
\end{table}

\begin{table}[H]
	\caption{Simulation results for sample size 8,000 \label{chap3:sim_res2}}
	\tiny
	\hspace{-1.75cm}
	\begin{tabular}{|lc|cccc|cccc|cccc|}
	\hline
	 \multicolumn{2}{|c|}{Scenario 1} & \multicolumn{4}{c|}{Na\"{i}ve} & \multicolumn{4}{c|}{MSM} & \multicolumn{4}{c|}{Proposed}\\
	Parameter & True value & Bias & RMSE & 95\% Coverage & AIL & Bias & RMSE & 95\% Coverage & AIL & Bias & RMSE & 95\% Coverage & AIL \\ 
	\hline
	$\Delta_{1,0}$ & -1.497 & -0.001 & 0.023 & 94.2 & 0.087 & 0.0003 & 0.023 & 94.0 & 0.087 & 0.0003 & 0.023 & 96.2 & 0.100 \\
	$\Delta_{01,00}$ & -1.499 & -0.002 & 0.036 & 95.3 & 0.143 & -0.001 & 0.036 & 94.8 & 0.143 & -0.001 & 0.037 & 95.5 & 0.151 \\
	$\Delta_{11,00}$ & -1.005 & -0.002 & 0.034 & 94.8 & 0.134 & -0.0007 & 0.034 & 95.1 & 0.134 & -0.001 & 0.035 & 98.7 & 0.183 \\
	$\Delta_{11,01}$ & 0.493 & 0.001 & 0.034 & 94.6 & 0.134 & 0.002 & 0.034 & 94.4 & 0.134 & 0.002 & 0.035 & 98.6 & 0.184 \\
	$\Delta_{001,000}$ & -1.502 & 0.005 & 0.057 & 95.0 & 0.222 & 0.005 & 0.057 & 99.0 & 0.289 & 0.005 & 0.058 & 95.4 & 0.235 \\
	$\Delta_{011,000}$ & -1.008 & 0.004 & 0.051 & 94.5 & 0.201 & 0.004 & 0.051 & 98.4 & 0.260 & 0.004 & 0.052 & 97.1 & 0.246 \\
	$\Delta_{111,000}$ & -0.504 & 0.005 & 0.051 & 95.1 & 0.201 & 0.007 & 0.052 & 98.3 & 0.261 & 0.006 & 0.053 & 99.7 & 0.369 \\
	$\Delta_{011,001}$ & 0.495 & -0.002 & 0.051 & 94.6 & 0.200 & -0.002 & 0.051 & 99.1 & 0.260 & -0.002 & 0.052 & 97.8 & 0.247 \\
	$\Delta_{111,001}$ & 1.000 & -0.001 & 0.052 & 94.2 & 0.201 & 0.0001 & 0.052 & 98.7 & 0.261 & -0.0005 & 0.054 & 99.8 & 0.369 \\
	$\Delta_{111,011}$ & 0.502 & 0.003 & 0.046 & 93.8 & 0.177 & 0.004 & 0.047 & 98.4 & 0.229 & 0.004 & 0.048 & 99.8 & 0.308 \\
	\hline
	 \multicolumn{2}{|c|}{Scenario 2} & \multicolumn{4}{c|}{Na\"{i}ve} & \multicolumn{4}{c|}{MSM} & \multicolumn{4}{c|}{Proposed}\\
	Parameter & True value & Bias & RMSE & 95\% Coverage & AIL & Bias & RMSE & 95\% Coverage & AIL & Bias & RMSE & 95\% Coverage & AIL \\ 
	\hline
	$\Delta_{1,0}$ & -3.367 & -0.047 & 0.055 & 59.6 & 0.109 & 0.002 & 0.029 & 94.0 & 0.113 & 0.003 & 0.029 & 95.9 & 0.125 \\
	$\Delta_{01,00}$ & -1.727 & -0.036 & 0.045 & 73.4 & 0.105 & -0.031 & 0.041 & 78.8 & 0.106 & -0.001 & 0.026 & 96.1 & 0.113 \\
	$\Delta_{11,00}$ & -1.199 & -0.134 & 0.139 & 4.0 & 0.142 & -0.018 & 0.041 & 92.4 & 0.144 & -0.001 & 0.036 & 96.9 & 0.161 \\
	$\Delta_{11,01}$ & 0.528 & -0.098 & 0.105 & 21.9 & 0.140 & 0.013 & 0.038 & 94.0 & 0.142 & -0.001 & 0.036 & 97.4 & 0.158 \\
	$\Delta_{001,000}$ & -1.727 & -0.029 & 0.049 & 87.9 & 0.156 & -0.024 & 0.047 & 93.2 & 0.170 & 0.0001 & 0.038 & 96.2 & 0.160 \\
	$\Delta_{011,000}$ & -1.183 & -0.066 & 0.075 & 54.3 & 0.141 & -0.048 & 0.060 & 77.7 & 0.153 & -0.001 & 0.036 & 97.7 & 0.156 \\
	$\Delta_{111,000}$ & -1.169 & -0.166 & 0.173 & 7.8 & 0.193 & -0.040 & 0.065 & 90.4 & 0.215 & -0.003 & 0.049 & 98.9 & 0.246 \\
	$\Delta_{011,001}$ & 0.544 & -0.038 & 0.050 & 81.5 & 0.131 & -0.025 & 0.042 & 92.1 & 0.142 & -0.002 & 0.032 & 97.2 & 0.145 \\
	$\Delta_{111,001}$ & 0.558 & -0.137 & 0.145 & 17.3 & 0.186 & -0.017 & 0.050 & 96.7 & 0.208 & -0.005 & 0.046 & 98.7 & 0.233 \\
	$\Delta_{111,011}$ & 0.013 & -0.098 & 0.108 & 38.0 & 0.174 & 0.010 & 0.046 & 96.4 & 0.194 & 0.0004 & 0.044 & 98.2 & 0.210 \\
	\hline
	 \multicolumn{2}{|c|}{Scenario 3} & \multicolumn{4}{c|}{Na\"{i}ve} & \multicolumn{4}{c|}{MSM} & \multicolumn{4}{c|}{Proposed}\\
	Parameter & True value & Bias & RMSE & 95\% Coverage & AIL & Bias & RMSE & 95\% Coverage & AIL & Bias & RMSE & 95\% Coverage & AIL \\ 
	\hline
	$\Delta_{1,0}$ & -2.347 & -0.123 & 0.126 & 1.5 & 0.113 & 0.002 & 0.029 & 94.5 & 0.113 & 0.003 & 0.029 & 95.7 & 0.124 \\
	$\Delta_{01,00}$ & -2.559 & -0.114 & 0.118 & 3.4 & 0.117 & -0.060 & 0.067 & 49.3 & 0.113 & -0.001 & 0.028 & 96.1 & 0.115 \\
	$\Delta_{11,00}$ & -3.062 & -0.230 & 0.234 & 0.1 & 0.164 & -0.032 & 0.052 & 86.5 & 0.159 & -0.004 & 0.040 & 97.3 & 0.183 \\
	$\Delta_{11,01}$ & -0.502 & -0.118 & 0.125 & 19.4 & 0.164 & 0.026 & 0.048 & 91.0 & 0.160 & -0.003 & 0.040 & 98.2 & 0.184 \\
	$\Delta_{001,000}$ & -2.820 & -0.125 & 0.133 & 16.6 & 0.171 & -0.063 & 0.076 & 78.3 & 0.192 & -0.002 & 0.039 & 95.6 & 0.157 \\
	$\Delta_{011,000}$ & -3.605 & -0.143 & 0.147 & 2.2 & 0.140 & -0.087 & 0.093 & 40.8 & 0.159 & -0.007 & 0.032 & 96.7 & 0.142 \\
	$\Delta_{111,000}$ & -4.032 & -0.290 & 0.296 & 0.1 & 0.225 & -0.081 & 0.099 & 81.5 & 0.265 & -0.010 & 0.057 & 98.7 & 0.282 \\
	$\Delta_{011,001}$ & -0.785 & -0.018 & 0.044 & 93.4 & 0.159 & -0.024 & 0.047 & 94.7 & 0.181 & -0.005 & 0.037 & 97.0 & 0.161 \\
	$\Delta_{111,001}$ & -1.217 & -0.160 & 0.171 & 22.4 & 0.238 & -0.013 & 0.062 & 97.1 & 0.278 & -0.008 & 0.059 & 98.7 & 0.311 \\
	$\Delta_{111,011}$ & -0.432 & -0.142 & 0.152 & 26.4 & 0.216 & 0.011 & 0.056 & 97.9 & 0.255 & -0.006 & 0.052 & 98.7 & 0.264 \\
	\hline
	\end{tabular}
\end{table}

\begin{table}[H]
	\caption{Descriptive statistics of 1996 Health and Retirement Study (baseline), part 1 \label{chap3:des_stat1}}
	\tiny
	\centering
	\begin{tabular}{|l|c|c|c|}
	\hline
		& No wealth shock & Ever wealth shock & \\
		Variables & Mean/Frequency (S.E./\%) & Mean/Frequency (S.E./\%) & $p$-value\\
	\hline
		Eventually survived?:	&					&					&	0.57	\\
$\quad$ Yes	& 6,207 (94.7) & 516 (94.0) & \\
$\quad$ No	& 350 (5.3) & 33 (6.0) &	\\
Cognitive score	& 17.07 (4.07) & 16.26 (4.35) &	$<0.01$	\\
BMI	& 27.21 (4.84) & 27.73 (5.40) & 0.03	\\
Self-reported health	&					&					&	$<0.01$	\\
$\quad$ Excellent	& 1,207 (19.9) &	83 (15.7)	&		\\
$\quad$ Very Good	& 2,126 (35.0)	& 128 (24.3) &		\\
$\quad$ Good	&	1,715	(28.2)	&	163	(30.9)	&		\\
$\quad$ Fair	&	763	(12.6)	&	103	(19.5)	&		\\
$\quad$ Poor	&	261	(4.3)	&	50	(9.5)	&		\\
Current Smoking status:	&					&					&	$<0.01$	\\
$\quad$ Never	&	2,353	(40.0)	&	166	(32.4)	&		\\
$\quad$ Former	&	2,410	(41.0)	&	187	(36.5)	&		\\
$\quad$ Current	&	1,116	(19.0)	&	159	(31.1)	&		\\
Alcohol consumption:	&					&					&	$<0.01$	\\
$\quad$ Never	&	3,799	(62.9)	&	347	(66.1)	&		\\
$\quad$ Moderate	&	1,686	(27.9)	&	116	(22.1)	&		\\
$\quad$ Heavy	&	555	(9.2)	&	62	(11.8)	&		\\
Insured?:	&					&					&	$<0.01$	\\
$\quad$ No	&	1,014	(15.5)	&	120	(21.9)	&		\\
$\quad$ Yes	&	5,543	(84.5)	&	429	(78.1)	&		\\
	Depression?:	&					&					&	$<0.01$	\\
$\quad$ No	&	4,922	(85.5)	&	361	(73.1)	&		\\
$\quad$ Yes	&	832	(14.5)	&	133	(26.9)	&		\\
Income (log transformed)	&	10.48	(1.21)	&	10.18	(1.45)	&	$<0.01$	\\
Labor force status:	&					&					&	$<0.01$	\\
$\quad$ Working	&	3,111	(51.2)	&	314	(59.6)	&		\\
$\quad$ Unemployed	&	96	(1.6)	&	13	(2.5)	&		\\
$\quad$ Retired	&	2,178	(35.9)	&	104	(19.7)	&		\\
$\quad$ Disabled	&	143	(2.4)	&	43	(8.2)	&		\\
$\quad$ Not in labor force	&	547	(9.0)	&	53	(10.1)	&		\\
Martial status:	&					&					&	$<0.01$	\\
$\quad$ Married	&	4,897	(80.8)	&	373	(70.8)	&		\\
$\quad$ Divorced	&	591	(9.7)	&	90	(17.1)	&		\\
$\quad$ Widowed	&	426	(7.0)	&	42	(8.0)	&		\\
$\quad$ Never Married	&	149	(2.5)	&	22	(4.2)	&		\\
Wealth rank in tertiles:	&					&					&	$<0.01$	\\
$\quad$ 0	&	1,728	(26.4)	&	326	(59.4)	&		\\
$\quad$ 1	&	2,360	(36.0)	&	124	(22.6)	&		\\
$\quad$ 2	&	2,469	(37.7)	&	99	(18.0)	&		\\
Gender:	&					&					&	0.08	\\
$\quad$ Male	&	3,113	(47.5)	&	239	(43.5)	&		\\
$\quad$ Female	&	3,444	(52.5)	&	310	(56.5)	&		\\
	\hline
	\end{tabular}
\end{table}

\begin{table}[H]
	\caption{Descriptive statistics of 1996 Health and Retirement Study (baseline), part 2 \label{chap3:des_stat2}}
	\scriptsize
	\centering
	\begin{tabular}{|l|c|c|c|}
	\hline
		& No wealth shock & Ever wealth shock & \\
		Variables & Mean/Frequency (S.E./\%) & Mean/Frequency (S.E./\%) & $p$-value\\
	\hline
Ever had diabetes?:	&					&					&	$<0.01$	\\
$\quad$ No	&	5,474	(90.2)	&	451	(85.6)	&		\\
$\quad$ Yes	&	596	(9.8)	&	76	(14.4)	&		\\
Ever had heart problems?:	&					&					&	0.43	\\
$\quad$ No	&	5,343	(88.0)	&	457	(86.7)	&		\\
$\quad$ Yes	&	730	(12.0)	&	70	(13.3)	&		\\
Ever had HBP?:	&					&					&	0.07	\\
$\quad$ No	&	3,888	(64.0)	&	316	(60.0)	&		\\
$\quad$ Yes	&	2,183	(36.0)	&	211	(40.0)	&		\\
Ever had psych problems?:	&					&					&	$<0.01$	\\
$\quad$ No	&	5,691	(93.7)	&	469	(89.2)	&		\\
$\quad$ Yes	&	380	(6.3)	&	57	(10.8)	&		\\
Ever had stroke?:	&					&					&	0.1	\\
$\quad$ No	&	5,912	(97.3)	&	506	(96.0)	&		\\
$\quad$ Yes	&	161	(2.7)	&	21	(4.0)	&		\\
Age	&	59.73	(3.19)	&	57.26	(2.18)	&	$<0.01$	\\
Number of education years centered	&	0.52	(2.93)	&	-0.17	(3.32)	&	$<0.01$	\\
Race:	&					&					&	$<0.01$	\\
$\quad$ Non-hispanic White	&	5,236	(79.9)	&	342	(62.3)	&		\\
$\quad$ Non-hispanic Black	&	759	(11.6)	&	120	(21.9)	&		\\
$\quad$ Hispanic	&	449	(6.8)	&	70	(12.8)	&		\\
$\quad$ Other	&	113	(1.7)	&	17	(3.1)	&		\\
	\hline
	\end{tabular}
\end{table}

\begin{table}[H]
	\caption{Change in unadjusted cognitive score between consecutive waves stratified by negative wealth shock status\label{chap3:cog_score_ts}}
	\centering
	\begin{tabular}{|l|ccc|}
		\hline
		& Never shocked & Ever shocked & Change in proportion shocked \\
		\hline
		Baseline to Wave 1 & 0.19 & -1.61 & 3.5\%\\
		Wave 1 to Wave 2 & -0.55 & 0.06 & 2.1\%\\
		Wave 2 to Wave 3 & -0.05 & -0.10 & 1.3\%\\
		\hline
	\end{tabular}
\end{table}

\begin{landscape}
\begin{table}
	\caption{Effect estimate of negative wealth shock on cognitive score for late middle aged adults in original Health Retirment Study cohort from 1996 to 2002. \label{chap3:anal_res}}
	%\small
	\begin{center}
	\begin{tabular}{|l|cc|cc|cc|cc|}
	\hline
	 & \multicolumn{2}{c|}{Na\"{i}ve} & \multicolumn{2}{c|}{Baseline adjusted$^\dagger$}& \multicolumn{2}{c|}{MSM*} & \multicolumn{2}{c|}{Proposed*} \\
	 & Estimate & 95\% CI & Estimate & 95\% CI & Estimate & 95\% CI & Estimate & 95\% CI\\
	 \hline
		2 years vs. no shock	&	-0.51	&	(-1.45,	0.35)	&	-0.51	&	(-1.37,	0.3)	&	-0.01	&	(-1.18,	1.07)	&	-0.13	&	(-0.83,	0.58)	\\
4 years vs. no shock	&	-0.69	&	(-1.45,	0.05)	&	-0.7	&	(-1.4, 0.03)	&	-0.31	&	(-1.23,	0.58)	&	0.18	&	(-0.73, 1.09)	\\
6 years vs. no shock	&	-1.95	&	(-2.62,	-1.25)	&	-1.94	&	(-2.6, -1.26)	&	-0.12	&	(-1.12,	0.89)	&	-0.18	&	(-0.87, 0.51)	\\
4 years vs. 2 years	&	-0.18	&	(-1.33,	1.04)	&	-0.19	&	(-1.26,	0.94)	&	-0.3	&	(-1.78,	1.15)	&	0.31	&	(-0.58,	1.20)	\\
6 years vs. 2 years	&	-1.45	&	(-2.54,	-0.38)	&	-1.43	&	(-2.46,	-0.4)	&	-0.1	&	(-1.61,	1.36)	&	-0.03	&	(-0.83, 0.78)	\\
6 years vs. 4 years	&	-1.26	&	(-2.27,	-0.2)	&	-1.24	&	(-2.2, -0.24)	&	0.19	&	(-1.11,	1.61)	&	-0.38	&	(-1.36,	0.61)	\\
	\hline
	\end{tabular}
	\end{center}
	{\footnotesize *Adjusted by gender, education category, race, cognitive score, BMI, self-reported health status, alcohol consumption, insurance status, depression status, income, labor force status, marital status, age, smoking status, diabetes status, heart condition, HBP status, psychological problem status, and stroke status at baseline as well as time-varying self-reported health status, alcohol consumption, insurance status, income, labor force status, smoking status, number of health conditions, and depression.
	
	$\dagger$Adjusted by gender, education category, race, cognitive score, BMI, self-reported health status, alcohol consumption, insurance status, depression status, income, labor force status, marital status, age, smoking status, diabetes status, heart condition, HBP status, psychological problem status, and stroke status at baseline.}
\end{table}
\end{landscape}

\end{document}